\documentclass[final,5p,times,twocolumn]{elsarticle} 
\usepackage{stfloats}
\usepackage{subfig}
\usepackage{colortbl}
\usepackage{graphicx,epsfig,color}
\usepackage{multirow}
\usepackage{amssymb,amsmath,array,wasysym}
\usepackage{upgreek}

\journal{Journal of \LaTeX\ Templates}

\begin{document}

\begin{frontmatter}

\title{Recent results of the technological prototypes of the CALICE  highly granular calorimeters}

\author{Roman P\"oschl}
\address{Laboratoire de l'Acc\'el\'erateur Lin\'eaire, Centre Scientifique d'Orsay, B\^at. 200, F-91898 Orsay}
\ead{poeschl@lal.in2p3.fr}

\author{On behalf of the CALICE Collaboration}



\begin{abstract}
The CALICE Collaboration is conducting R\&D for highly granular calorimeters with an emphasis on detectors for Linear Colliders. This contribution briefly summarises recent tests of large scale technological prototypes of a silicon tungsten electromagnetic calorimeter and hadron calorimeters featuring either a gaseous medium or scintillator with a SiPM readout as active material. 
\end{abstract}


\end{frontmatter}


\section{Introduction}

The design of particle detectors at future high-energy physics experiments and, in particular, at linear colliders is oriented towards the usage of Particle Flow Algorithms (PFA) for the event reconstruction. These algorithms aim to achieve good jet energy resolution of the order of 3-4\% for jet energies between around 45\,GeV and several 100\,GeV. The algorithms reconstruct individual particles by combining signals in tracking systems and in high granularity calorimeters~\cite{Brient:2002gh, Morgunov:2002pe, Thomson:2009rp}. 

The primary objective of the CALICE Collaboration is the development, construction and testing of highly granular hadronic and electromagnetic calorimeters for future particle physics experiments. The first stage of this effort is marked by the successful running of so-called physics prototypes that delivered the proof of principle that highly granular calorimeters can be  constructed and operated~\cite{Adloff:2012dla}. The main options in terms of absorber and active material are summarised in Table~\ref{tab:calicecals}.

\definecolor{light-gray}{gray}{0.70}
\begin{table*}[h]
\begin{center}
\begin{tabular}{@{} |rlccc| @{}}
  \hline
  Project                 & Purpose of prototype                                   & Absorber                     & Sensitive part         & Status     \\
  \hline
  \multirow{2}*{{\bf AHCAL}}    & Physics                        & Stainl.\,steel/Tungsten          & Scintillator           & Completed           \\
  \arrayrulecolor{light-gray} \cline{2-5} \arrayrulecolor{black} 
  ~                       & {\bf Technological }                 & {\bf Stainl.\,steel}                 & {\bf Scintillator}           & {\bf Ongoing}                \\
  \hline
  TCMT                    & Physics                       & Stainl. steel                & Scintillator           & Completed           \\
  \hline
  \multirow{2}*{DHCAL}    & \multirow{2}*{Physics \& Technological}       & \multirow{2}*{Stainl.\,steel/Tungsten} & RPC                    &  \multirow{2}*{Completed} \\
  ~                       &                                 &                              & Partially GEM          &                     \\ 
  \arrayrulecolor{light-gray} \hline \arrayrulecolor{black} 
  \multirow{2}*{{\bf SDHCAL}}   & \multirow{2}*{{\bf Physics \& Technological}} & \multirow{2}*{{\bf Stainl.\,steel}} & {\bf GRPC}                    & \multirow{2}*{{\bf Ongoing}} \\
  ~                       &                                          &                              & Partially $\upmu$Megas  &                     \\
  \hline
\multirow{2}*{{\bf SiW Ecal}}    & Physics                        & Tungsten               & Silicon          & Completed           \\
  \arrayrulecolor{light-gray} \cline{2-5} \arrayrulecolor{black} 
  ~                       & {\bf Technological}                 & {\bf Tungsten}                & {\bf Silicon}           & {\bf Ongoing}                \\
  \hline
\multirow{2}*{ScW Ecal}    & Physics                        & Tungsten               & Scintillator          & Completed           \\
  \arrayrulecolor{light-gray} \cline{2-5} \arrayrulecolor{black} 
  ~                       & Technological                  & Tungsten                & Scintillator           & Ongoing                \\
  \hline
\end{tabular}
\end{center}
\caption{\sl Overview of calorimeter prototypes developed and tested by CALICE. In bold face letters those projects that are subject of this article. Acronyms are defined in the text.}
\label{tab:calicecals}
\end{table*}

Since around 2011 the prototypes address technological and engineering questions coming along with granular calorimeters.

\section{Overview on technological prototypes}

The keywords that describe the current R\&D phase are; Realistic dimensions with structures of up to 3m; Integrated front end electronics and compact digital electronics for readout;
Small power consumption by using power pulsed electronics exploiting the beam structure of linear colliders. The beam is delivered in bunch trains. In between the bunch trains
the bias currents can be shut down.

\subsection{Silicon tungsten electromagnetic calorimeter - SiW ECAL}

A stack of the technological prototype of the SiW ECAL featuring seven layers with 1024 cells of $5\times 5\,{\rm mm^2}$ cell size each has been tested in beam tests at DESY and at CERN. In general the test comprised different layouts of PCBs and silicon wafer thicknessesof about (325\,$\upmu {\rm m}$) and 650\,$\upmu {\rm m}$. Since 2013 the different versions of the prototype have been operated in power pulsed mode. 

\begin{figure*}[ht]
\centering
\includegraphics[width=0.4\linewidth]{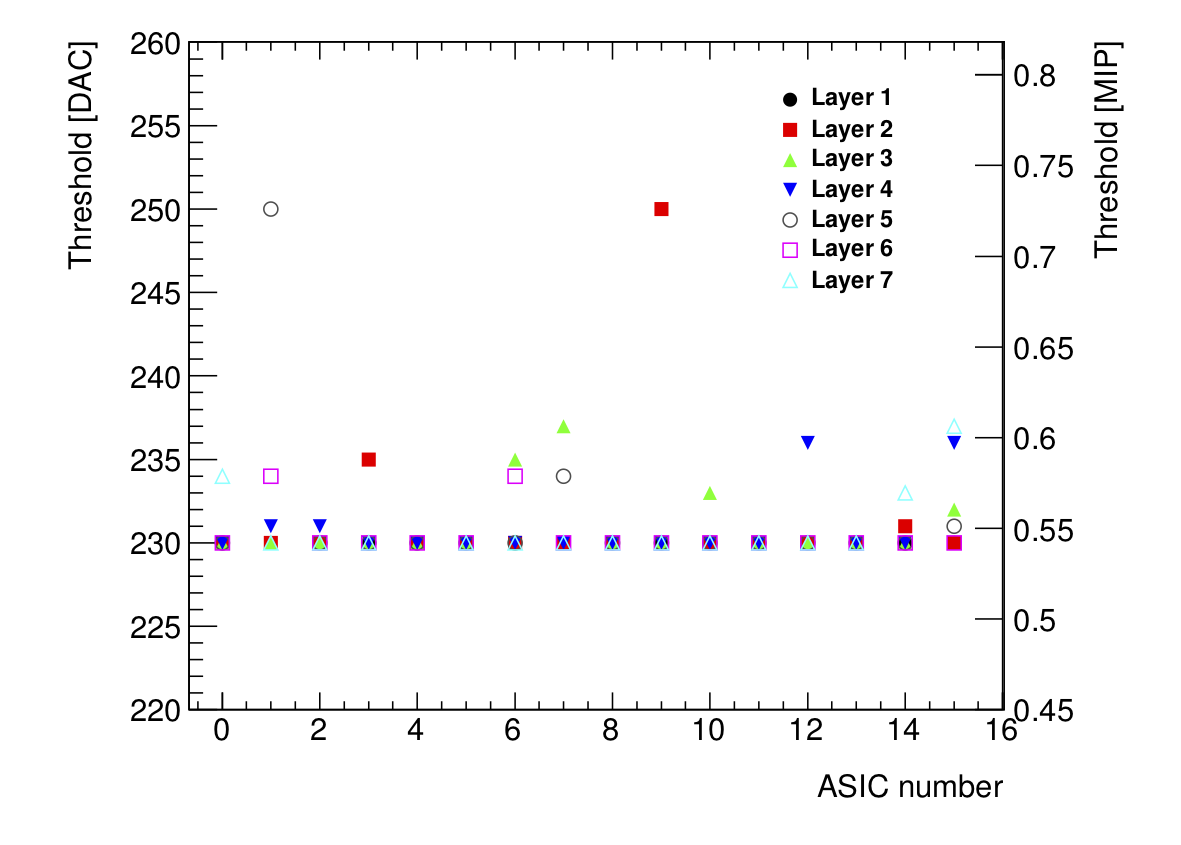}
\hspace{1.5cm}
\includegraphics[width=0.4\linewidth]{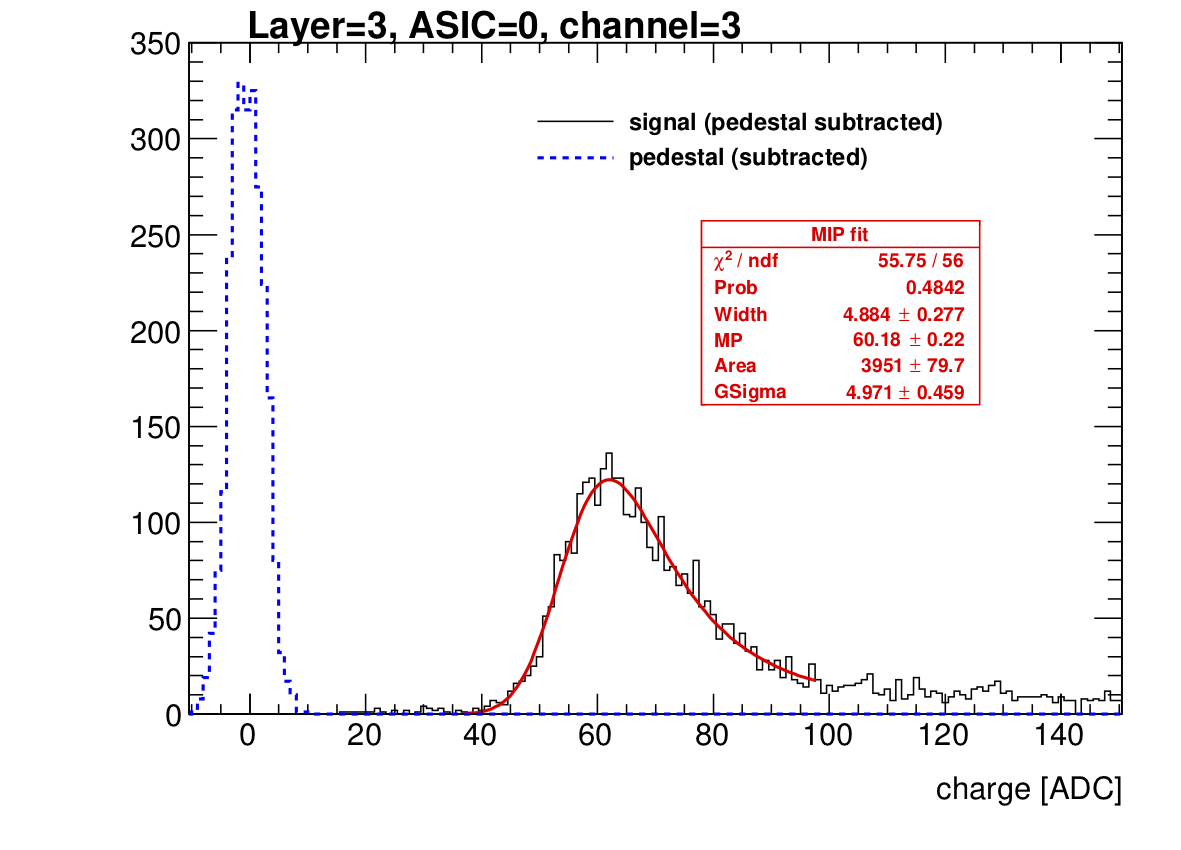}
\caption{\label{fig:ecal-thresh-mip} \sl \underline{Left:} On-ASIC trigger thresholds for different layers of the SiW ECAL stack. \underline{Right:}  Typical MIP and noise spectrum recorded in SiW ECAL cells. Results are taken from~\cite{Bilokin:2018gfn}.}
\end{figure*}

A central goal is to demonstrate the feasibility of  the required signal-over-noise ratio above 10 as expected from the technological choice. As all ASICs used in the technological prototypes the  SKIROC2 ASICs is auto-triggered. It features therefore a branch for triggering and one for actual charge readout, called `ADC-Branch'. It is important to know the signal-over-noise ratio in both branches. For 325$\upmu {\rm m}$ wafer thickness a signal-over-noise ration of $12.9\pm3.4$ is reported for the trigger branch~\cite{Bilokin:2018gfn}. This allows for setting uniformly the trigger threshold at a level of about 1/2 MIP as shown in the left part of Fig.~\ref{fig:ecal-thresh-mip}. The right part shows a typical MIP spectrum after digitisation. From this one can conclude that the actual signal-over-noise ratio in the ADC-Branch is about 20. The detector is regularly operated in power pulsed mode. An individual layer has been tested in a magnetic field of 1\,T with no apparent degradation of performance~\cite{Bilokin:2018gfn}.

\subsection{The analogue hadronic calorimeter - AHCAL}

\begin{figure*}[ht]
\centering
\includegraphics[width=0.44\linewidth]{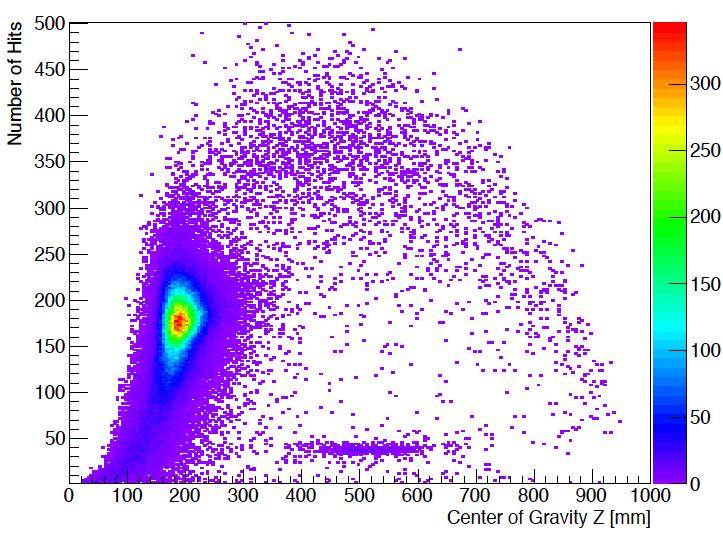}
\hspace{1.5cm}
\includegraphics[width=0.46\linewidth]{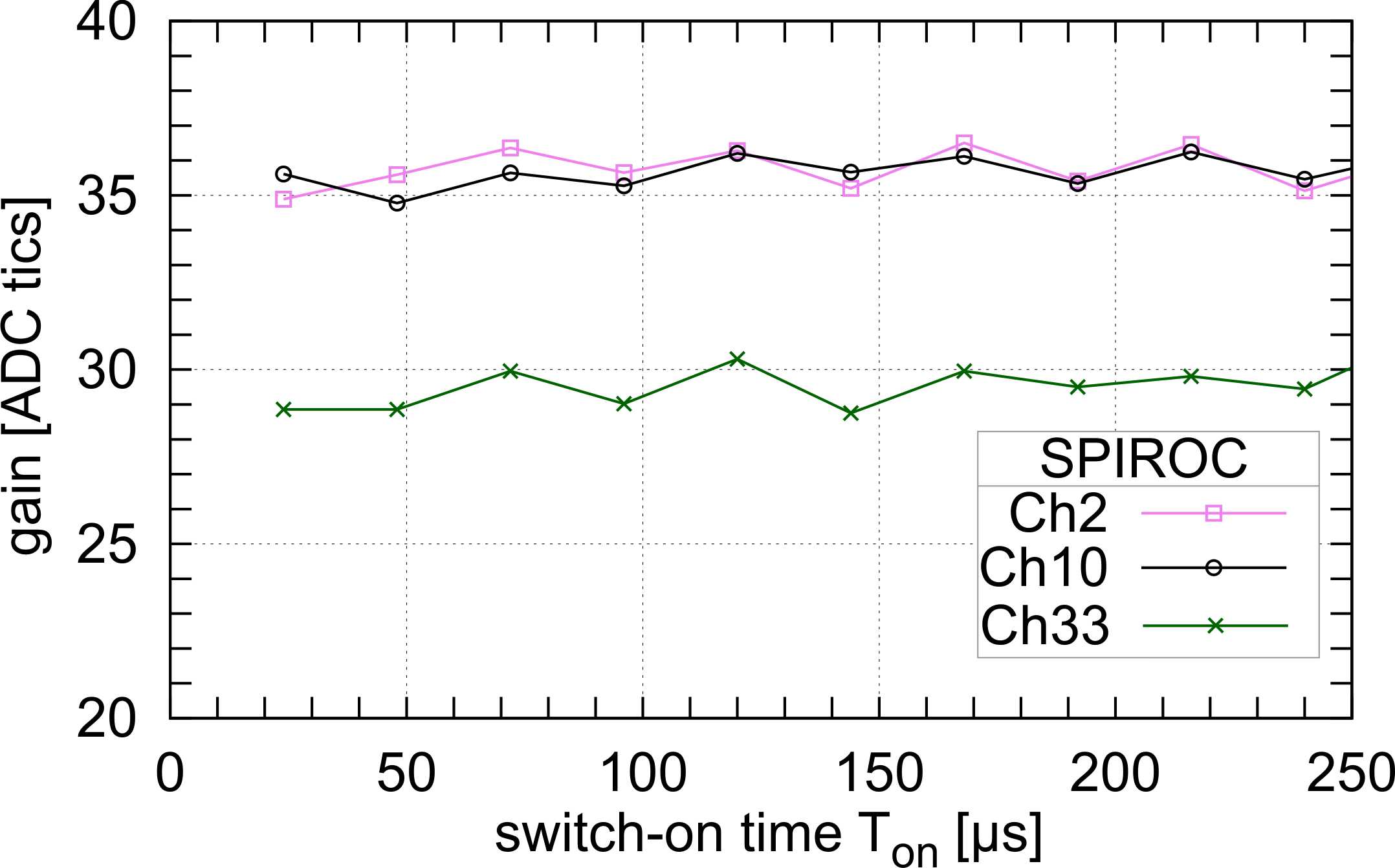}
\caption{\label{fig:ahcal-pid} \sl \underline{Left:} Frequency of number of hits as a function of the layer number in the CALICE AHCAL. \underline{Right:} Gain of the SPIROC2 ASICs for different switch on times $T_{on}$ in power pulsing mode.}
\end{figure*}

The active medium of the AHCAL is scintillating tiles that are readout by silicon-photomultipliers.
In 2018 a stack of the AHCAL with 39 active layers sandwiched between 1.7\,cm steel absorber plates ($4\,\lambda_I$ in total) was commissioned and tested with hadron beams at CERN.
An active layer has a transverse size of about $1\,{\rm m^2}$ divided into readout cells of $3\times3\,{\rm cm^2}$.
The setup was completed by a tail catcher comprising 12 layers with 5.4\,cm steel absorber  ($4\,\lambda_I$).
The left part of Figure~\ref{fig:ahcal-pid} demonstrates that different particles lead to different patterns in terms of number of hits as function of their longitudinal position.

Two further things that were tested is power pulsing and also the timing resolution. A critical parameter is the switch-on time $T_{on}$, which is the time needed until the electronics is settled after the enabing of the bias currents in power pulsing mode. The AHCAL is currently operated with $T_{on}=150\,{\rm \upmu s}$
Single photon electron spectra of silicon photomultipliers allow for monitoring the gain of the ASIC channels.
The right part of Figure~\ref{fig:ahcal-pid} shows the gain as a function of $T_{on}$. The gain is the same for all $T_{on}$ but in particular there is no difference in gain for $T_{on}=150\,{\rm \upmu s}$ compared with the gain at $T_{on}>150\,{\rm \upmu s}$. For further details see~\cite{Reinecke:2016pfj}. 

A time resolution of 5\,ns is observed for a clock frequency of 250\,kHz used in the TDC ramp of the SPIROC ASIC~\cite{Brianne:2018rur}. The ASICs allows for running with a 5 MHz clock. It is therefore likely that the design goal of 1\,ns time resolution can be reached.

\subsection{The semi-digital hadronic calorimeter - SDHCAL \label{sec:sdhcal}}

The SDHCAL is a high granularity sampling calorimeter with 48 Glass Resistive Plate Chambers (GRPC) used as active media with a transversal size of $1\,{\rm m^2}$ divided into $96 \times 96$ readout cells of $1\,{\rm cm^2}$ each. Absorber layers are made of 2\,cm thick stainless steel plates. The total depth is about $6\,\lambda_I$. For further details see~\cite{Baulieu:2015pfa}. With a smaller version of the calorimeter power pulsing has been tested in a magnetic field of 3\,T and no degradation of performance has been observed~\cite{Caponetto:2011ep}. 

\begin{figure*}
\centering
\includegraphics[width=0.31\linewidth]{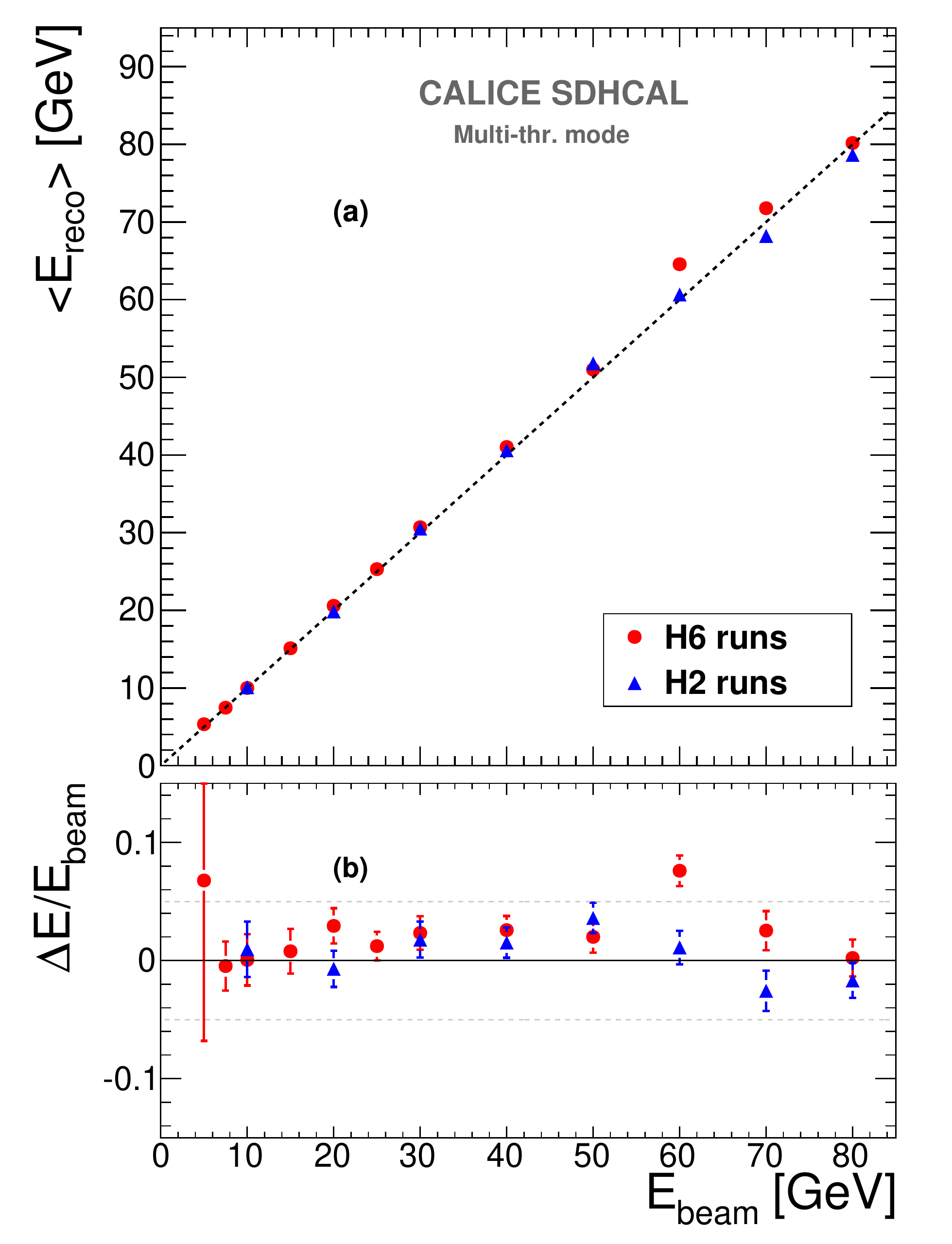}
\hspace{1.5cm}
\includegraphics[width=0.42\linewidth]{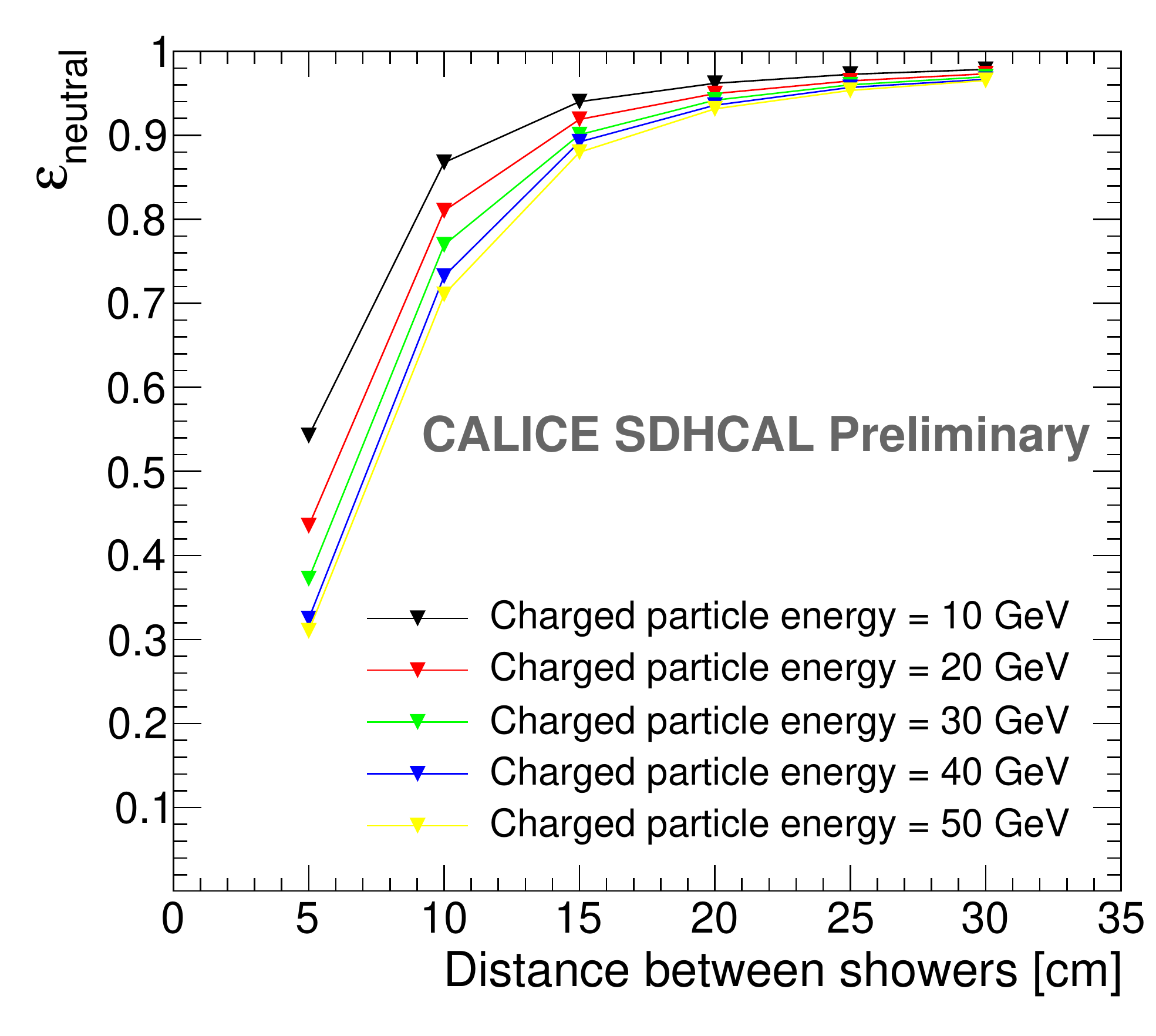}
\caption{\label{fig:sdhcal-bit-res} \sl \underline{Left:} Linearity of the CALICE SDHCAL in ternary threshold mode. \underline{Right:} Efficiency of the separation of a 10\,GeV neutral hadron from a charged hadron for different energies of the charged hadron.}
\end{figure*}

The avalanche generated in GRPC layers by an interacting particle can be readout in one-bit (binary mode) of a two-bit (ternary mode) resolution. In case of the  ternary mode the number of hits above a given threshold are multiplied with weighting factors obtained from a $\chi^2$ analysis of the reconstructed energy~\cite{Buridon:2016ill}.  
The left part of Fig.~\ref{fig:sdhcal-bit-res} shows linearity of the SDHCAL reponse for two different beam test campaigns. The response is linear within $\sim$5\% and the good agreement between the two data sets demonstrates the stability of the detector operation. 

The ultimate goal of a jet energy resolution over the entire energy range covered by linear colliders, depends critically on the particle separation capabilities of the granular calorimeters.  The right part of Figure~\ref{fig:sdhcal-bit-res} shows the separation of a 10\,GeV neutral particle from a charged particle for different energies of the charged particle~\cite{CALICE:2015xn} using an adpated version of the ARBOR PFA~\cite{Ruan:2014paa}. Here the separation is estimated from the calorimeter hits that are assigned to the corresponding particle after reconstruction. This example demonstrates that starting at around a distance of 10\,cm, the two particles can be unambiguously separated.  
 
\section{Different schemes of hadronic energy reconstruction}

\begin{figure*}
\centering
\includegraphics[width=0.4\linewidth]{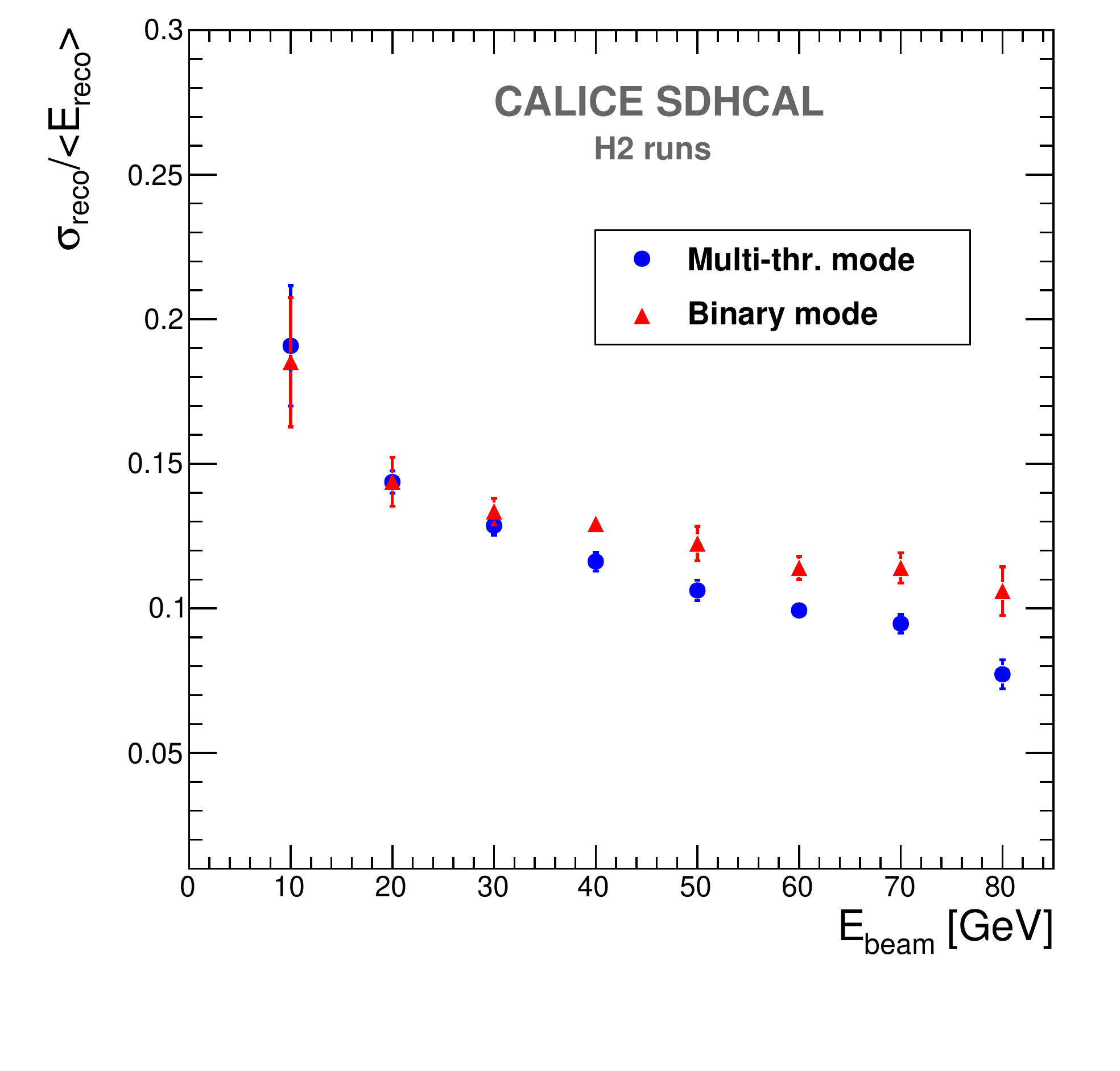}
\hspace{1.5cm}
\includegraphics[width=0.35\linewidth]{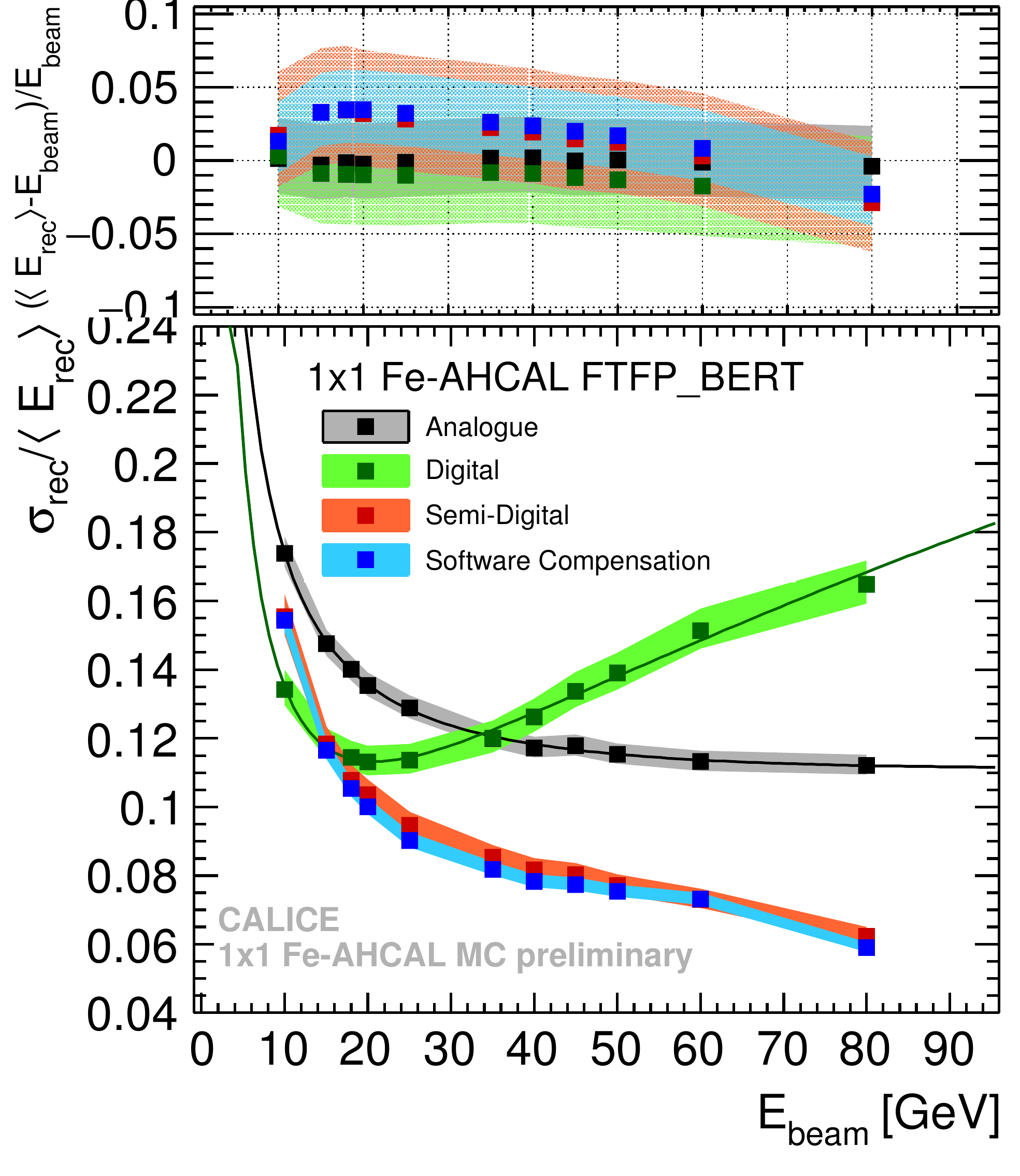}
\caption{\label{fig:reco-schemes} \sl \underline{Left:} Energy resolution of the CALICE SDHCAL for binary and ternary threshold modes. \underline{Right:} Energy resolution of the AHCAL for different reconstruction techniques from a simulation study using $1\times 1\,{\rm cm^2}$ cell size.}
\end{figure*}

As already sketched in Sec.~\ref{sec:sdhcal} the CALICE Calorimeters use different schemes for hadronic energy reconstruction that depend on the readout scheme. The left part of Fig.~\ref{fig:reco-schemes} compares the energy resolutions obtained for the binary mode and ternary of the SDHCAL. The additional thresholds improve the resolution in particular at higher energies. The analogue readout of the AHCAL allows for applying software compensation schemes in which the deposited energy in a cell is weighted to e.g. ensure an $e/h$ ratio of about one. 
For the simulation study in Ref.~\cite{CALICE:2014xk} different readout technologies were emulated for an AHCAL of $1\times 1\,{\rm cm^2}$ cell size. The plot on the right hand side of Fig.~\ref{fig:reco-schemes} shows that a semi-digital reconstruction and the analogue reconstruction using software compensation yield the best results in terms of energy resolution.

\section{Summary and outlook}

This article presents a short summary on the status of the technological prototypes of the CALICE Collaboration. Prototypes of a silicon tungsten electromagnetic calorimeter, a analogue scintillating-tiles steel hadronic calorimeter and a semi-digital calorimeter with GRPC as sensitive material and steel absorber have been examined in beam tests in recent years. The examples given in this article show that key performances can be achieved. These are triggering and reading out of signals as small as 1/2 MIP, stable operation in power pulsed mode and timing resolution at the ns level. The hadronic energy resolution of both hadron calorimeter prototypes is of the order of 5-10\% at highest energies. Different reconstruction methods are used for the different prototypes depending on their readout scheme. All of them yield  software weighting methods for the optimisation of the energy reconstruction and resolution supporting the final goal of the 3\% to 4\% jet energy resolution of PFA based event reconstruction as for example demonstrated in Ref.~\cite{Tran:2017tgr}.

The maturity of the individual prototypes permits to setup common beam tests comprising different CALICE technological prototypes as e.g. comprising the SiW ECAL and SDHCAL in 2018 or even prototypes with other projects as e.g. the common beam tests between the CMS-HGCAL and the CALICE AHCAL is 2017 and 2018.  
In contrast to the operation and linear electron positron colliders granular calorimeters at hadron colliders or circular electron-positron colliders will require active cooling and therefore a revision of the detector layout. 
Timing resolution of granular calorimeters becomes increasingly important. At the HL-LHC timing is essential to mitigate the harsh pile-up conditions. Intuitively timing should also also help to identify different components of a hadronic shower and therefore support the association of shower particles to primary particles. An excellent timing could make a granular calorimeter suitable for time-of-flight measurements. Excellent timing puts however strict requirements on the front-end-electronics entailing e.g.\,a higher power consumption. Solutions for granular calorimeters without active cooling will have to be developed in the coming years.

\section*{Acknowledgements}
On behalf of CALICE the author would like to thank the organisers of the VCI2019 Conference for the opportunity to present the CALICE R\&D program. CALICE thanks DESY and CERN for their support and for the reliable and efficient beam operation. We acknowledge funding from the European Union's Horizon 2020 Research and Innovation program under Grant Agreement no. 654168.

\section*{References}
\begin{tiny}
\bibliography{mybibfile}
\end{tiny}
\end{document}